\documentclass[a4paper]{eccomas_paper-2024}
\usepackage{graphicx}

\usepackage{hyperref}

\usepackage{graphicx}
\usepackage{amsmath}
\usepackage{amsfonts}
\usepackage{amssymb}
\usepackage{xfrac}

\usepackage{mathtools}

\usepackage[numbers]{natbib}

\usepackage[labelformat=simple]{subcaption}   	
\usepackage{cleveref} 
\crefformat{figure}{Figure~#2#1#3} 
\Crefformat{figure}{Figure~#2#1#3} 
\crefformat{equation}{equation~#2#1#3} 
\Crefformat{equation}{Equation~#2#1#3} 

\usepackage{tikz}
\usepackage{pgfplots}
\usepackage{xargs}
\usepackage{ifthen}
\usepackage{standalone}

\usepackage{physics}
\newcommand{\MainPath}{..}
\newcommand\pt[1]{\boldsymbol{#1}}
\def\TrimCurve{\pt{C}}

\def\tikzOff{}
\input{mycommands.tex}

\graphicspath{{.}}	

\title{REDUCING MESHING REQUIREMENTS FOR ELECTROSTATIC PROBLEMS USING A GALERKIN BOUNDARY ELEMENT METHOD}

\author{BENJAMIN~MARUSSIG$^1$, THOMAS~R\"{U}BERG$^2$, J\"{U}RGEN~ZECHNER$^2$, LARS~KIELHORN$^2$ and THOMAS-PETER~FRIES$^3$}

\heading{Benjamin~Marussig, Thomas~R\"{u}berg, J\"{u}rgen~Zechner, Lars Kielhorn and Thomas-Peter~Fries}

\address{$^{1}$Institute of Applied Mechanics, \\
Graz Center of Computational Engineering (GCCE), \\
Graz University of Technology\\
Technikerstaße 4/II, 8010 Graz, Austria\\
www.mech.tugraz.at, www.gcce.tugraz.at\\
e-mail: marussig@tugraz.at
\and
$^{2}$TailSiT GmbH\\
Nikolaiplatz 4, 8020 Graz, Austria\\
www.tailsit.com
\and
$^{3}$Institute of Structural Analysis, \\
Graz Center of Computational Engineering (GCCE), \\
Graz University of Technology\\
Lessingstraße 25/II, 8010 Graz, Austria\\
www.ifb.tugraz.at, www.gcce.tugraz.at
}


\abstract{%
    This work focuses on model preparation for electrostatic simulations of CAD designs to realize a rapid virtual prototyping concept. We present a boundary element method (BEM) allowing discontinuous fields between surfaces. 
    The corresponding edges of the CAD model are enhanced with the data required to integrate over non-conforming elements.
    Finally, we generate a mesh for each CAD surface.
    The approach is verified via numerical experiments and shows excellent agreement with conforming BEM results. 
 }

\begin{document}
\thispagestyle{empty}

\section{INTRODUCTION}

\begin{figure}
    \centering
    \begin{subfigure}[b]{0.55\textwidth}    
        \includegraphics[trim={17cm 3.8cm 13cm 5cm},clip,width=0.8\textwidth]{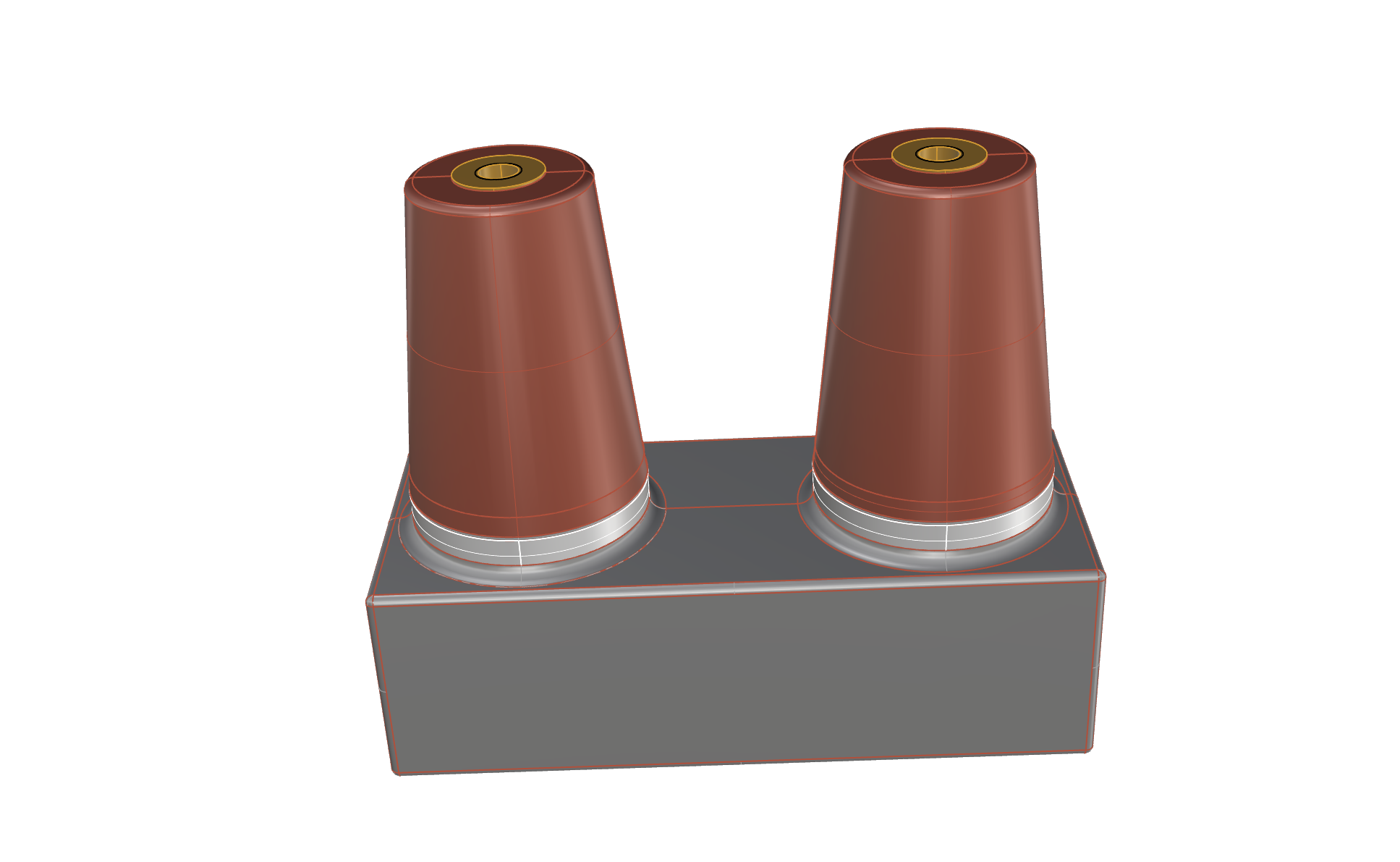}\\
        \hspace*{0.6\textwidth}\includegraphics[width=0.18\textwidth]{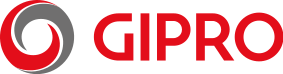}
    \subcaption{Insulator design}
    \label{fig:EpoxyU-Verbinder}    
    \end{subfigure}
    \begin{subfigure}[b]{0.33\textwidth}
        \reflectbox{\includegraphics[width=0.85\textwidth]{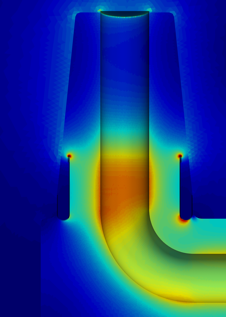}}
        \subcaption{Electric field}
        \label{fig:EpoxyU-EField}
    \end{subfigure}
    \caption{Virtual prototyping example: (a) CAD model of the designed device and (b) the related finite element solution of the electric field in the right cone for virtual design assessment.}
    \label{fig:EpoxyU-VerbinderAndEField}
\end{figure}

We aim to develop a simulation tool for rapid virtual prototyping for electrostatics problems. 
In particular, the focus lies on the computation of electric fields of CAD designs of electric devices.
For instance, \cref{fig:EpoxyU-Verbinder} illustrates the CAD model of an insulator from our industry partner GIPRO, for which the interior electric field, as shown in \cref{fig:EpoxyU-EField}, is computed to assess weaknesses in the design.
The physical quantities of interest are commonly obtained by using finite element methods (FEM).

\begin{figure}
    \centering
    \includegraphics[width=0.35\textwidth]{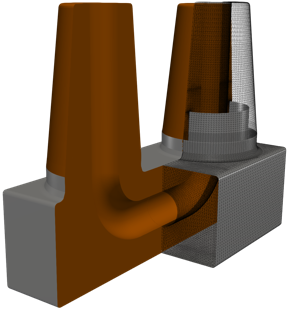}
    \caption{Interior of \cref{fig:EpoxyU-Verbinder}, where the right grid indicates the volumetric mesh for FEM simulations.}
    \label{fig:FEMMesh}
\end{figure}

\begin{figure}[t]
    \centering
    \begin{subfigure}[b]{0.48\textwidth}    
        \includegraphics[trim={17cm 8cm 13cm 5cm},clip,width=0.9\textwidth]{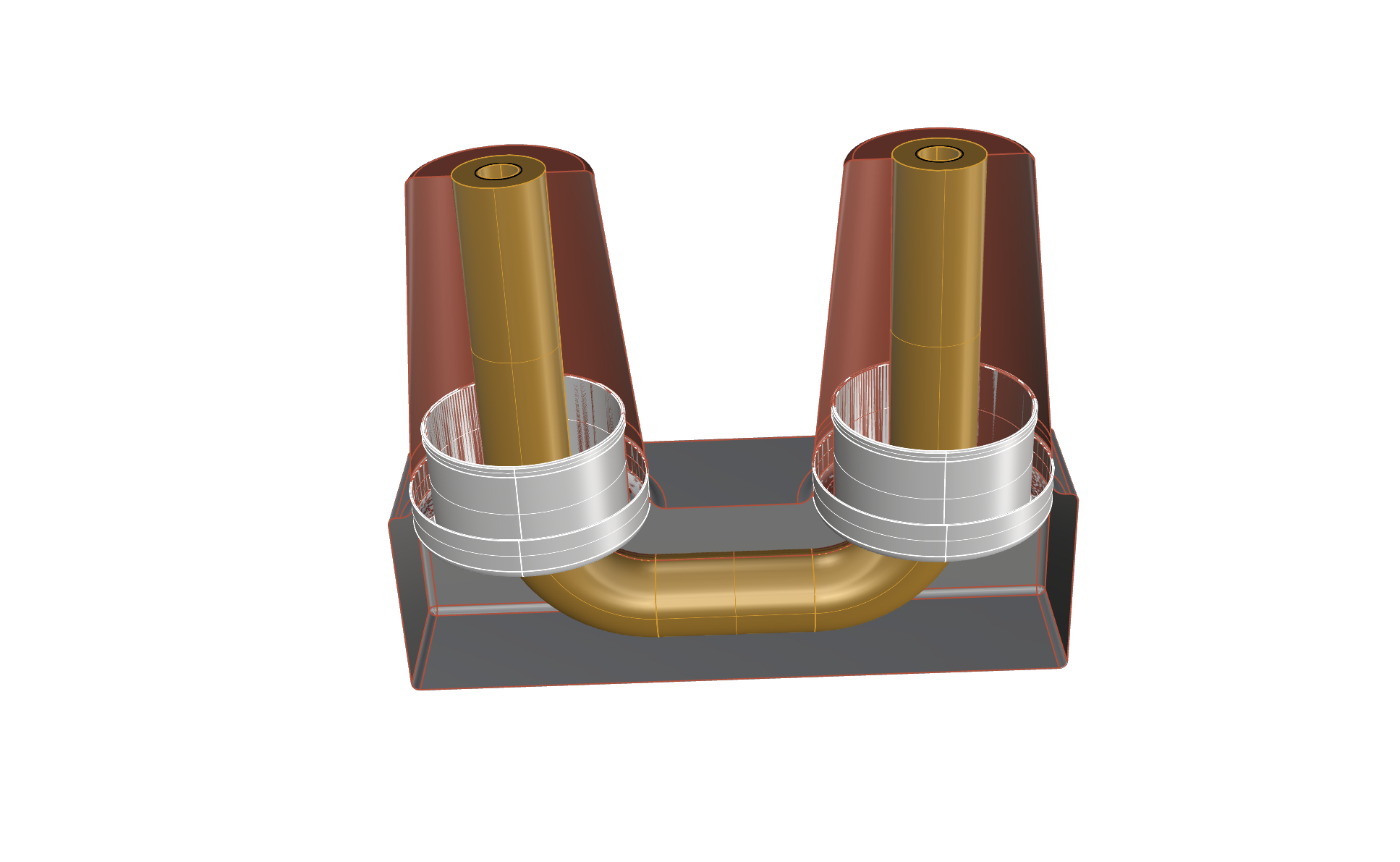}
        \subcaption{Graphics display}        
    \end{subfigure}
    \begin{subfigure}[b]{0.48\textwidth}
        \includegraphics[trim={17cm 8cm 13cm 5cm},clip,width=0.9\textwidth]{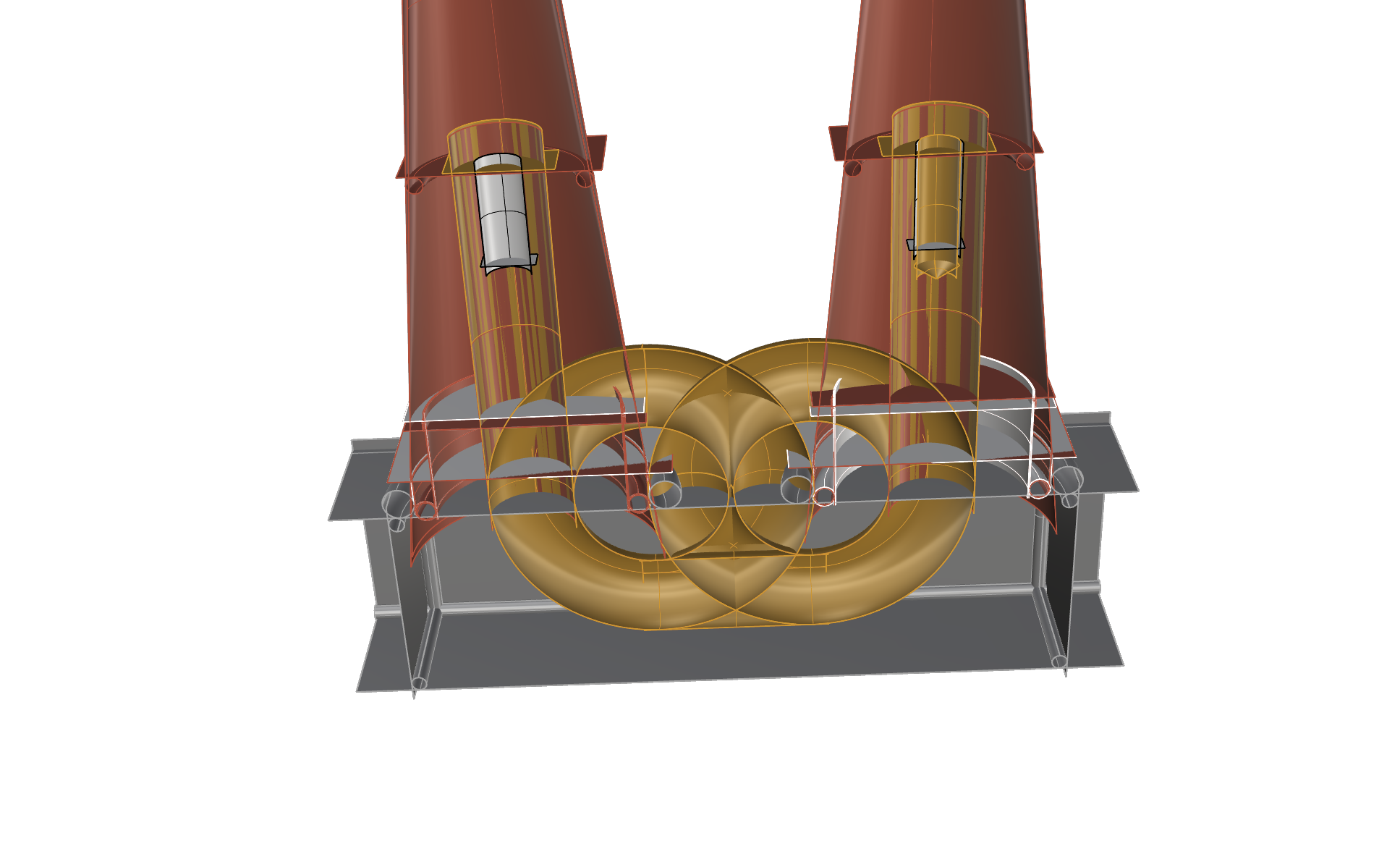}
        \subcaption{Geometric representation}
    \end{subfigure}
    \caption{Trimmed CAD models: (a) cut through the model as shown to a user and (b) the mathematical representation of its surfaces, which is the basis for the meshing process.}
    \label{fig:Timming}
\end{figure}

But before FEM can be applied,
two major challenges have to be addressed: First, CAD models are based on a boundary representation (b-rep), while FEM requires a spatial discretization of the objects' volume.
\cref{fig:FEMMesh} indicates the interior volume of the insulator in \cref{fig:EpoxyU-Verbinder}. 
This volume has to be represented by a volumetric FEM-mesh.
Moreover, the exterior domain has to be meshed as well to account for the infinite extent of the fields.
Second, even the CAD b-reps are usually not ready for FEM simulations since they consist of trimmed surfaces. 
The problem here is that the parametrization of such surfaces does not coincide with the visualized object, as illustrated in \cref{fig:Timming}.
These challenges manifest themselves as artifacts (inconsistent orientations, large-scale overlap, small gaps and overlaps, singular vertices; see \cite{botsch2010polygon} for details) during the meshing process.
In the best case, these artifacts can be addressed by semi-automatic schemes. Hence, they are an enduring and time-consuming problem for virtual prototyping.  

This work presents an approach to reduce the extensive meshing efforts typically required in electrostatic simulations and foster closer interaction with CAD designs and electrostatic simulations. 
In particular, we propose a Boundary Element Method (BEM) combined with enhanced edge data from trimmed surfaces.
The overall approach circumvents the need for volumetric meshing and allows for non-conforming surface meshes,  which reduces the meshing efforts to mesh each trimmed surface independently.

\Cref{sec:BEM} introduces the BEM formulation, while \cref{sec:CAD} details the data enhancement of the CAD model. In \cref{sec:results}, numerical experiments demonstrate the overall virtual prototyping capabilities.

\section{BOUNDARY ELEMENT METHOD}
\label{sec:BEM}
\begin{figure}
  \centering
    \includegraphics[width=0.9\textwidth]{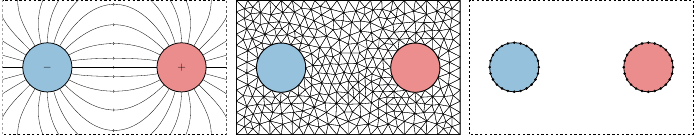}
    \caption{Electric field between two electrodes (left), its truncated discretization for the analysis with the FEM (middle) and the surface discretization required for the simulation with the BEM.}            
    \label{fig:BEMvsFEMMesh}
\end{figure}

For the conventional numerical analysis of electromagnetic fields with the FEM, the surrounding air has to be incorporated into the analysis model. %
Its discretization may become complex and time-consuming. %
Typically, the mesh has to be truncated at artificial boundaries introducing additional modeling errors due to the infinite extent of electromagnetic fields.
That can be completely avoided by using a BEM, for which a discretization of the boundary is sufficient.
In this work, we target the simulation of epoxy resin insulators and bushings where the dielectric material is considered to be linear.
Thus, we utilize the BEM for the simulation of the interior of the electric device too.
This results in a simulation method that requires only a surface description for the whole model.
The comparison in \cref{fig:BEMvsFEMMesh} clearly shows the reduced meshing effort by choosing BEM for the simulation.

Typically, conforming surface meshes are required for analysis.
That means that neighboring elements share a complete edge and all the nodes on it.
Therefore, the creation of analysis-suitable meshes is more difficult.
To reduce that effort, we choose a Galerkin-based BEM formulation that relieves the requirements for the mesh in such a way that non-conformal surface elements are permitted at least along the b-rep boundaries and trimming lines. 



\subsection{ELECTROSTATICS}
The electrostatic field equations result from Maxwell's equation by removing all time derivatives:
\begin{align}
  \curl \mathbf{E} = 0, \; \div \mathbf{D} = \rho,\; \curl \mathbf{H} = 0,\; \div \mathbf{B} = 0
  \label{eq:Maxwell}
\end{align}
In this case, the electric field $\mathbf{E}$ and electric displacement $\mathbf{D}$ are decoupled from the magnetic field strength $\mathbf{H}$ and flux density $\mathbf{B}$, which are not considered any further.
For the solution of \cref{eq:Maxwell}, we use the scalar potential $u$ with $\mathbf{E} = -\grad u$ and the material law $\mathbf{D} = \varepsilon \mathbf{E} = -\varepsilon_r \varepsilon_0 \mathbf{E}$ to obtain the Poisson equation
\begin{align}
  -\div \varepsilon\grad u = \rho,
  \label{eq:poisson}
\end{align}
where $\varepsilon_0$ and $\varepsilon$ denote the electric permittivity of the vacuum and the considered dielectric material, respectively.
$\varepsilon_r=\varepsilon/\varepsilon_0$ denotes the dimensionless relative permittivity.

On the boundary between two regions, the electric transmission conditions have to be fulfilled, that are the tangentional continuity of $\mathbf{E}$ and the normal continuity of $\mathbf{D}$. 
For the scalar potential, this leads to
\begin{align}
  \gamma_D^+ u - \gamma_D^- u = 0 && \text{and} && \varepsilon_r^+ \gamma_N^+ u - \varepsilon_r^- \gamma_N^- u = 0,
\end{align}
where $\gamma_D$ denotes the Dirichlet trace and $\gamma_N$ the Neumann trace.
Furthermore, the $\sfrac{+}{-}$ superscripts refer to the limit $\delta\rightarrow0$ of the evaluation at $\vb{x}\pm\delta\vb{n}$, $\vb{x}\in\Gamma$ with $\Gamma=\partial\Omega$ and $\Omega\in \mathbb{R}^3$.
We consider that there are no volume charges such that $\rho=0$ and relate the surface charge density $\sigma = -\varepsilon \gamma_N u$ with the scalar potential on the surface.

For the design of a bushing or insulator, the potential $u$ on the electrodes is constant and given by the applied voltage $g_E$. 
The unknown quantities are the surface charge density $\sigma$ on the electrodes and the interfaces between dielectrics as well as the voltage $u$ on the dielectric interfaces. So-called floating potentials, which may be sheets, hold an unknown constant potential.
The total charge on floating potentials is zero \cite{Amann2014a}.

\subsection{BOUNDARY ELEMENT EQUATIONS}
\label{sec:BEMformulation}

The single layer potential
\begin{align}
  u(\vb{x}) &= \int_\Gamma U(\vb{x}, \vb{y}) \sigma(\vb{y}) \dd{s_y} = \left( S\sigma \right)(\vb{x})   && \vb{x} \notin \Gamma
  && \text{with} && U(\vb{x},\vb{y}) = \frac{1}{4\pi} \frac{1}{| \vb{x}-\vb{y} |} &&
\end{align}
fulfills \cref{eq:poisson} for unknown charge densities $\sigma$ defined on the domain's boundary $\Gamma$ and with $\rho=0$.
By applying the Dirichlet and Neumann traces, we end up with
\begin{align}
  \gamma_D^\pm (S\sigma) &= V(\sigma) = \int_\Gamma U(\vb{x},\vb{y}) \sigma(\vb{y}) \dd{s_y} 
  \shortintertext{and}
  \gamma_N^\pm (S\sigma) &= [\mp  \tfrac{1}{2}I + K^\prime] (\sigma) = \mp \tfrac{1}{2} \sigma + \int_\Gamma (\grad_x U(\vb{x},\vb{y})) \cdot \vb{n}(\vb{x}) \dd{s_y} 
\end{align}
where $V$ denotes the single layer and $K^\prime$ the adjoint double layer boundary integral operator.

For the approximation of the surface charges, we use discontinuous shape functions $\varphi^i$ spanning $L_2(\Gamma)$
\begin{align}
  \sigma(\vb{x}) \approx \sigma_h(\vb{x}) = \sum_{i=1}^I \varphi^i(\vb{x}) \sigma^i.
\end{align}
That particular choice allows the numerical approximation of the boundary integral operators by means of a Galerkin method without any requirement on the continuity across elements in the mesh. 

Splitting electrodes ($E$), floating potentials ($F$), dielectrics as well as the surrounding air ($D$) through surface patches $\Gamma_{a}$, discretizing the boundary integral operators by means of the above-chosen shape functions and the Galerkin method and by resorting with respect of the given boundary data, we define a block system of boundary integral equations 
\begin{align}
  \left[ 
  \begin{array}{cccc}
    \mp \tfrac{1}{2} M_{DD} + K_{DD} & K_{DE} & K_{DF} & 0 \\
    V_{ED} & V_{EE} & V_{EF} & 0 \\
    V_{FD} & V_{FE} & V_{FF} & -H_F  \\
    F_{D} & F_{E} & F_{F} & 0  
        \end{array}
  \right]
  \left[ 
  \begin{array}{c}
    \sigma_{D}\\
    \sigma_{E}\\
    \sigma_{F}\\
    \alpha  
  \end{array}
  \right]
  =
  \left[ 
  \begin{array}{c}
    0\\
    g_{E}\\
    0\\
    0 
  \end{array}
  \right].
  \label{eq:BlockSystem}
\end{align}
$M_{DD}$ is the mass matrix, whereas $F$ and $H$ are matrices that are derived from the zero-charge-condition on floating potentials and $\alpha$ is their unknown voltage.
The exact derivation of the above system of equations would exceed the scope of this paper. 
Further details can be found in \cite{Amann2014a} and the references therein.

\subsection{MATRIX ENTRIES AND GEOMETRY INFORMATION}

Considering the single layer operator for example, the matrix entries in \cref{eq:BlockSystem} are 
\begin{align}
  V_{ab}[i,j] &= \int_{\Gamma_a} \int_{\Gamma_b} \varphi^i(\vb{x}) U(\vb{x}, \vb{y} ) \varphi^j(\vb{y}) \dd{s_y} \dd{s_x}.
\end{align}
As $\vb{x} \rightarrow \vb{y}$, the integral kernel $U(\vb{x},\vb{y}) \rightarrow \infty$ becomes singular and special measures must be taken into account for the calculation of affected matrix entries in $K_{DD}$, $V_{EE}$, and $V_{FF}$.
Three different types of singularity are distinguished for the Galerkin method, depending on the pair of considered elements: 
1.)~identical elements, 2.)~elements sharing a common edge, and 3.)~elements sharing a common vertex \cite{sauter2010}.
Each case involves a different quadrature for the evaluation of the corresponding entry \cite{erichsen1998}.

The presented formulation allows a non-conforming mesh, thus elements need to be subdivided in order to match one of the three types. 
There is nothing left to do for case 1, but if non-conforming elements touch each other, the elements have to be subdivided with respect to the hanging nodes of the other element and then classified case as 2 or 3.
For the subdivision, the location of the hanging node is required and to be provided by the geometry information.


\section{CAD MODEL PREPARATION}
\label{sec:CAD}

\subsection{SIMULATION DOMAINS AND BOUNDARY CONDITIONS}

The initial step of virtual prototyping is the specification of the domains relevant to the simulation. We do that directly in the CAD system by associating the surfaces of the CAD model to groups representing a dielectric, an electrode, or a floating potential. 
Each group represents a subdomain of the overall problem and includes information about the corresponding boundary conditions.
However, before performing the BEM simulation, we need to enhance the surface data of each group, which is outlined in the subsequent sections.

\subsection{ENHANCED TRIMMED SURFACE DATA}

\begin{figure}[b!]        
    \centering
    \includegraphics[width=0.7\textwidth,trim=0 75 0 0, clip]{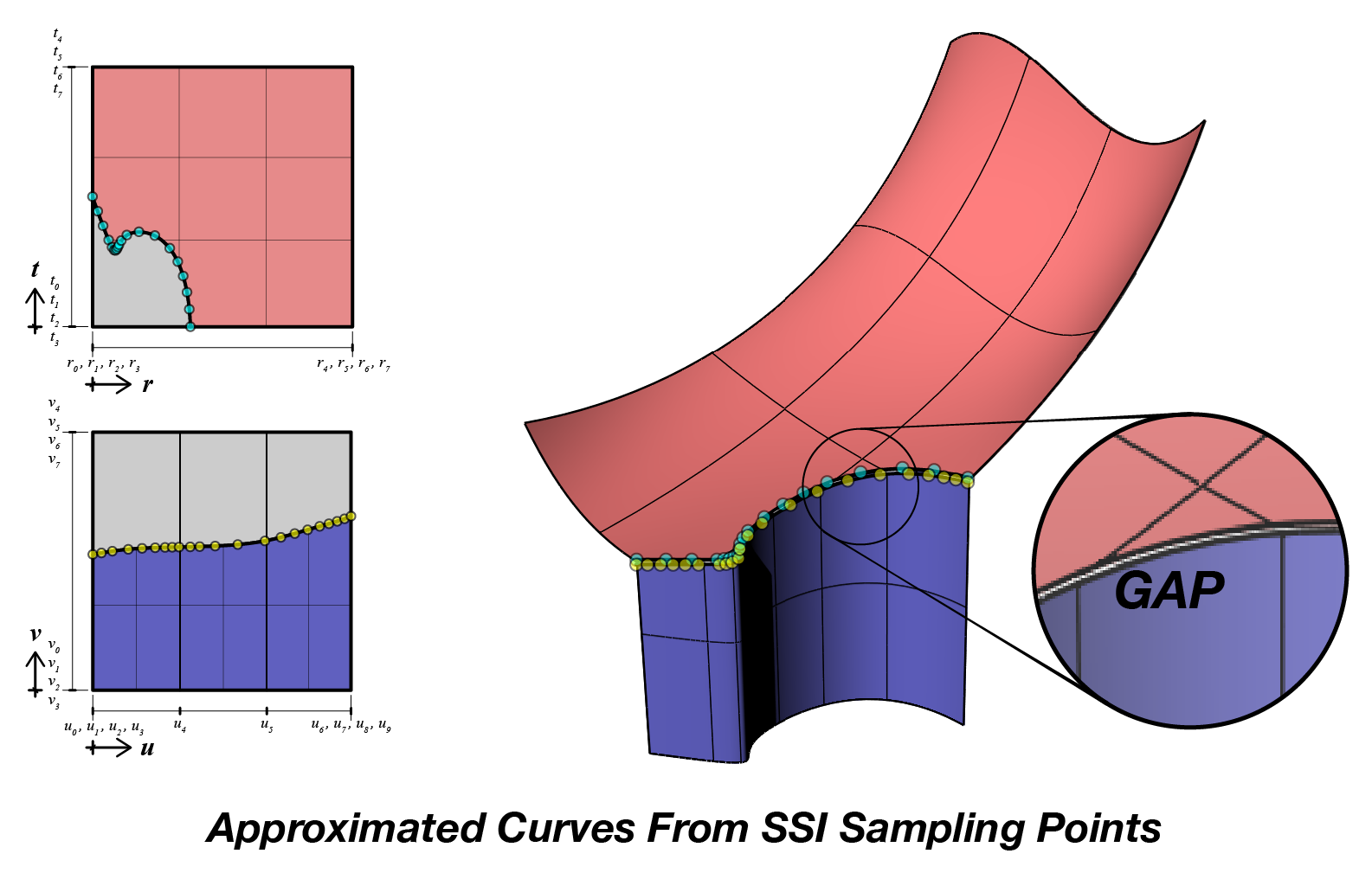}
    \caption{Surface-surface intersection (SSI) of two surfaces: (left) the restriction of the NURBS parameter spaces to (right) the visualized surfaces in the CAD display.}
    \label{fig:trimCAD}
\end{figure}

Non-uniform rational B-splines (NURBS) are the bread and butter technology in engineering CAD systems because of their high representation power. 
Trimmed surfaces are also omnipresent in such systems since they allow the application of NURBS to arbitrarily  shaped surfaces.
Consider a surface-surface intersection (SSI), as shown in \cref{fig:trimCAD}.
Instead of creating a new NURBS surface with a parameterization fitted to the object defined by both surfaces joined at their intersection, 
the display of the initial NURBS surfaces is only adjusted so that only a portion is visualized \cite{urick2017web}.
The remaining parts are \emph{trimmed} away, but the underlying mathematical representation of the surfaces remains unchanged.
%
While trimmed surfaces are an effective concept for model visualization, they are not suitable for simulations in general.
In fact, translating trimmed surfaces into analysis-suitable representations is an intensive research area, e.g.~\cite{Massarwi2018a,Wassermann2019a,Urick2020a,marussig2017a}. 

\begin{figure}
    
  \centering
  \def\tkzscale{0.8}
  \tikzsetnextfilename{VEGASurfaceEnhancement}
  \input{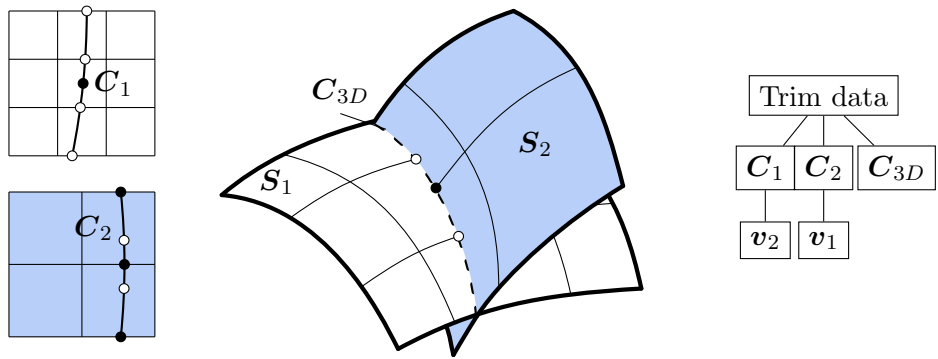}
  \caption{Data of trimmed surfaces: the initial surface, $\pt{S}_i$, the representations of the intersection curve in the related parameter spaces, $\TrimCurve_i$, and the model space, $\pt{C}_{3D}$. The color of the parameter spaces on  the left matches the color of the related surfaces shown in the middle. The isoline-vertex data, $\pt{v}_i$, is the enhancement required for the presented approach.}
  \label{fig:trimCADEnhanced}

\end{figure}

Thanks to the proposed BEM formulation, we can significantly reduce the effort for integrating trimmed surfaces into the simulation pipeline.
The starting point is the standard data of a trimmed surface. As indicated in \cref{fig:trimCADEnhanced}, it consists of the initial surface $\pt{S}$, the trim curve in the parameter space defining the visible area interface $\TrimCurve$, and a related representation of the trim curve in the model space ${C}_{3D}$. 
It is important to note that the different curves representing the intersection do not coincide. They are all independent approximations with no link between each other, as detailed in \cite{urick2017web,marussig2017a}.
We establish such a link by computing vertices, $\pt{v}_i$, that mark the closest points of one surface's isolines to the image of the other surface's $\TrimCurve$. 
In \cref{fig:trimCADEnhanced}, these points are displayed in black (in the top parameter space of $\pt{S}_1$) and white (in the bottom parameter space of $\pt{S}_2$).
The accumulation of these dots divides each $\TrimCurve_i$ into four segments, and their ends mark critical points of the own or intersection surface.
Hence, they enhance each trimmed surface with information about their neighbors and provide a connection between each other.

Robustness is of utmost importance for the usability of the enhancement process. 
Conceptually, the segments are generated within the CAD tool (Rhino 7) using its inherent routines, such as the provided closest point projection (CPP). 
Thus, the CAD model data is enhanced by utilizing the validation checks of the design tool and does not require any data translation. 
On top of that, we introduced a sequence of validations where segments are identified with several iterations of CPP to assess if the results converge to an unambiguous solution.

\subsection{NON-CONFORMING BEM MESHES}

\begin{figure}[th]
    \centering
    \begin{subfigure}[b]{0.48\textwidth}    
        \includegraphics[width=0.97\textwidth]{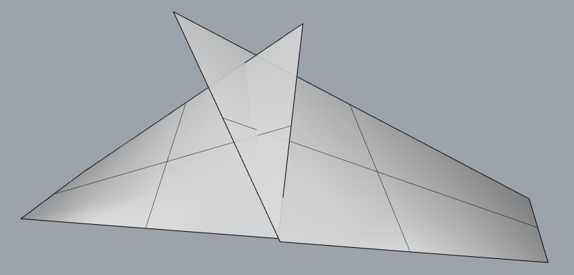}
        \subcaption{Initial surfaces}  
    \end{subfigure}
    \begin{subfigure}[b]{0.48\textwidth}
        \includegraphics[width=0.97\textwidth]{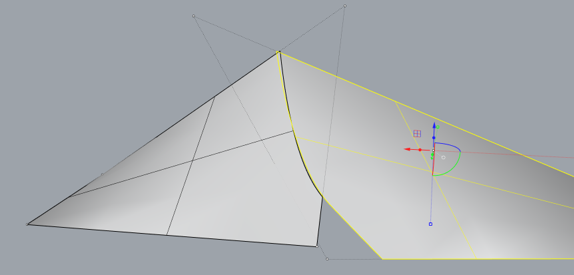}
        \subcaption{Trimmed CAD model}
    \end{subfigure}
    \begin{subfigure}[b]{0.48\textwidth}    
        \includegraphics[width=0.97\textwidth]{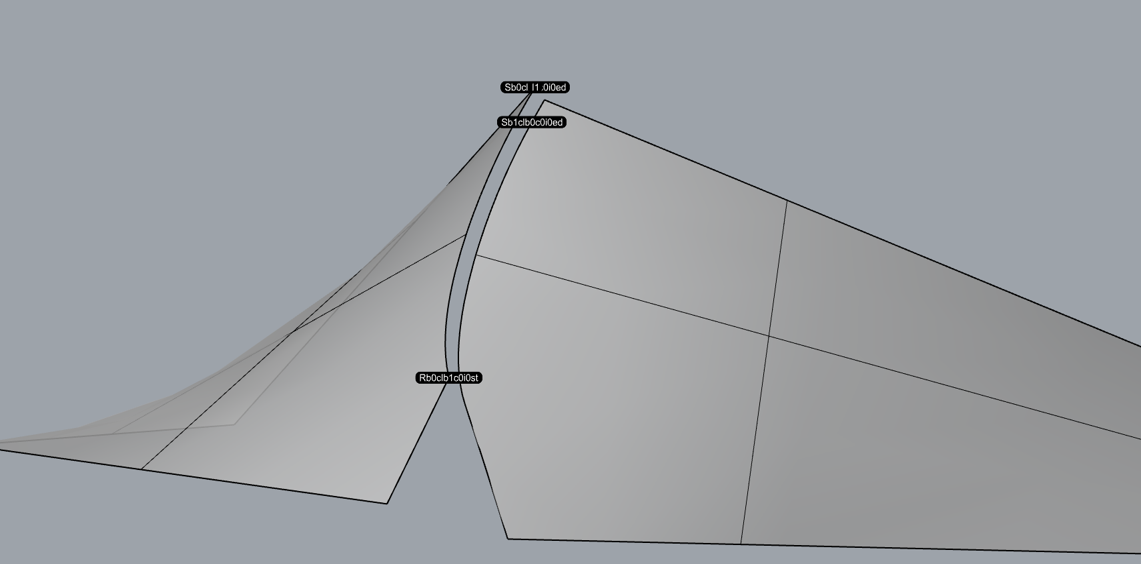}
        \subcaption{Trimmed CAD model enhancement}  
        \label{fig:meshenhancementEnhancedModel}
    \end{subfigure}
    \begin{subfigure}[b]{0.48\textwidth}
        \includegraphics[width=0.97\textwidth]{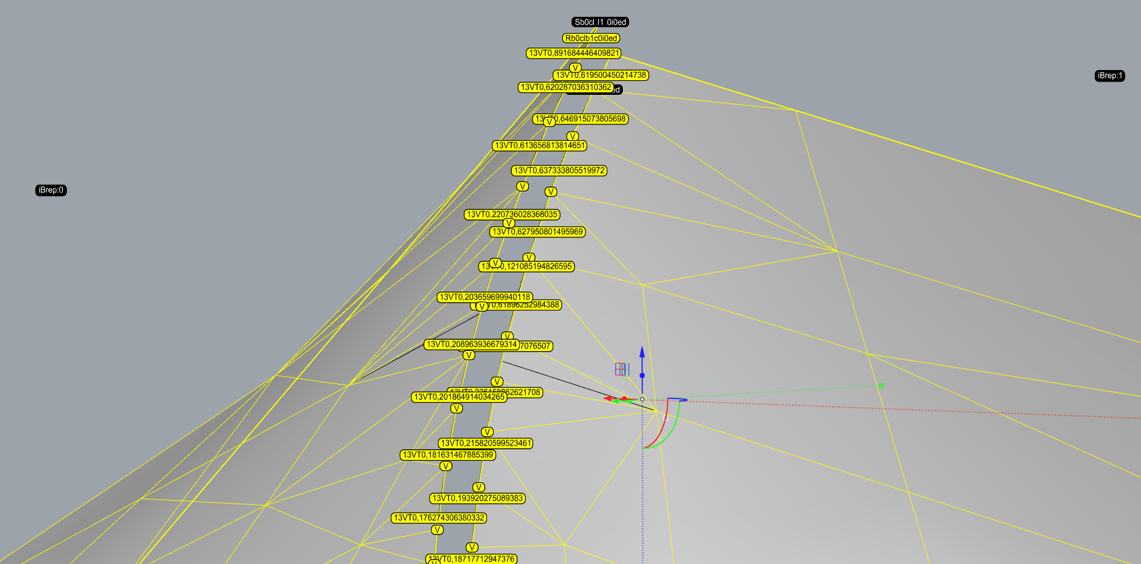}
        \subcaption{Enhanced non-conforming mesh}
        \label{fig:meshenhancementEnhancedMesh}
    \end{subfigure}
    \caption{Analysis data workflow: black bars in (c) indicate enhanced segment vertices, and the yellow bars in (d) indicate locations of mesh nodes in the elements of the neighboring mesh.}
    \label{fig:meshenhancement}
\end{figure}

\Cref{fig:meshenhancement} illustrates the overall workflow for preparing the BEM mesh.
The CAD model enhancement discussed in the previous subsection leads to \cref{fig:meshenhancementEnhancedModel}, where black bars along the intersection indicate the additional vertex data.
We emphasize that the black lines within the surfaces are no surface isolines but help to mark the location of the final close-up in \cref{fig:meshenhancementEnhancedMesh}.  

To obtain the final analysis model for virtual prototyping, we first mesh each trimmed surface separately from each other.
In other words, we make the meshing process as simple as possible.
In mesh generation, one distinguishes between unstructured mesh generation \cite{Owen_1998a,Si_2015a} and (block-)structured mesh generation \cite{Ali_2016a,Fogg_2015a,Bosnjak_2023a,Bosnjak_2024b}. Here, we prefer a hybrid approach where a structured mesh is used as a starting point, however, elements cut by the trimming curved are decomposed into (unstructured) sub-triangles.
Then, the vertices along the boundary of each mesh are mapped to the adjacent elements of the neighboring surfaces.
In \cref{fig:meshenhancementEnhancedMesh}, yellow bars indicate these locations.
This mapping process takes the enhanced trim curve segments and utilizes this data to link the position of a mesh vertex in one surface with the element of another surface mesh,
thereby providing the final missing link for the proposed BEM formulation.

The resulting mesh in the interior of a trimmed surface is conforming and non-conforming only along its boundaries.
The parameter space of non-conforming elements is enriched by the location of an adjacent mesh vertex from another surface, allowing their proper subdivision for the singular integration.

\section{NUMERICAL EXPERIMENTS}
\label{sec:results}

In this section, we demonstrate the performance of the proposed BEM formulation that uses the described non-conforming meshes with linear elements.
The first example verifies the implementation using a problem with an analytic solution. Then, we investigate the impact of the non-conformity using a practical example of an electric device and compare the results with a conforming simulation. 
%
In all examples, the mesh preparation steps are implemented as a plugin of the CAD software Rhino 7.

\subsection{SPHERICAL ELECTRODES}

\begin{figure}
    
  \centering
  \def\tkzscale{0.5}
  \tikzsetnextfilename{VEGATwoElectrodesSetUp}
  \input{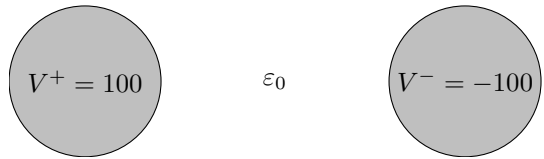}
  \caption{Boundary conditions of the first problem with the vacuum's permittivity $\epsilon_0$ and the voltages $V^+$ and $V^-$ of the two spherical electrodes.}
  \label{fig:electrodesBC}

\end{figure}

\begin{figure}
    
  \centering
  \def\tkzscale{0.33}
  \tikzsetnextfilename{VEGATwoElectrodes}
  \input{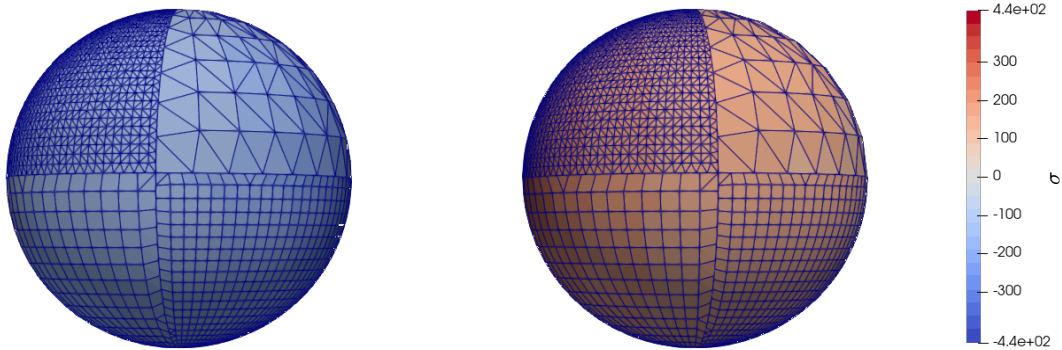}
  \caption{Non-conforming mesh and surface charge density of the spherical electrode problem.}
  \label{fig:electrodesSol}

\end{figure}

We consider two spherical electrodes in a vacuum with a voltage of $\pm 100$, respectively, as indicated in \cref{fig:electrodesBC}.  
The distance between the spheres is equal to their radius.  
Each sphere is modeled with eight trimmed surfaces, which are meshed using different element types and mesh sizes to enforce non-conformity along their boundaries. 
\Cref{fig:electrodesSol} illustrates the resulting discretization.  
Furthermore, the colors show the surface charge density $\sigma$ obtained by the BEM simulation.

To assess the quality of the results, we use the capacitance 
\begin{align}
C = \frac{Q}{V^+ - V^-} && \text{with} && Q(\Gamma) = \int_\Gamma {\sigma} ds
\end{align}
where $Q(\Gamma)$ is the total surface charge on the electrodes.
For the displayed discretization, we obtain a numerical capacitance of $9.612$, while the analytic capacitance is $9.567$. Thus, the relative error is $4.7 \cdot 10^{-3}$.

\subsection{HIGH VOLTAGE BUSHING}

\begin{figure}
    
  \centering
  \def\tkzscale{0.22}
  \tikzsetnextfilename{VEGABushingSetUp}
  \input{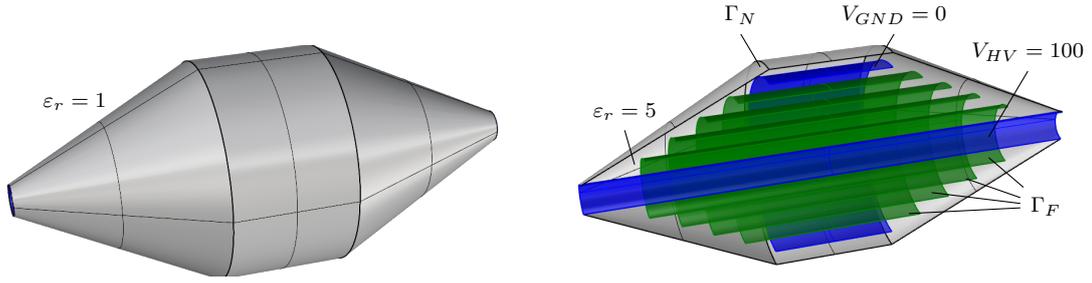}
  \caption{Bushing model}
  \label{fig:bushingBC}

\end{figure}
This example considers the design of a high-voltage bushing shown in \cref{fig:bushingBC}.
At the center of the model is a cylindrical conductor with radius $r$ and voltage $V_{HV}=100$.
It is embedded in a solid dielectric material with the relative permittivity $\varepsilon_r=5$, while for the exterior domain $\varepsilon_r=1$.
The related boundary $\Gamma_N$ is represented by two truncated cones (radii $r$ and $7r$, height $12r$) and a cylinder (radius $7r$, height $8r$).
Inside the solid material, five metallic cylindrical foils surround the conductor. The distance between them is $r$, and their height is chosen such that the distance to $\Gamma_N$ in the longitudinal direction is also $r$. 
The outermost foil is grounded; thus, the voltage is prescribed as $V_{GND}=0$.
The other foils are floating potentials $\Gamma_F$ that shall enforce a uniform potential distribution along the bushing's surface.

We set up two BEM meshes for this model: the first utilizes a conforming quadrilateral mesh, whereas the second re-meshes the cylinder of $\Gamma_N$ with coarse triangular elements.
\begin{figure}
    
  \centering
  \def\tkzscale{0.26}
  \tikzsetnextfilename{VEGABushing}
  \input{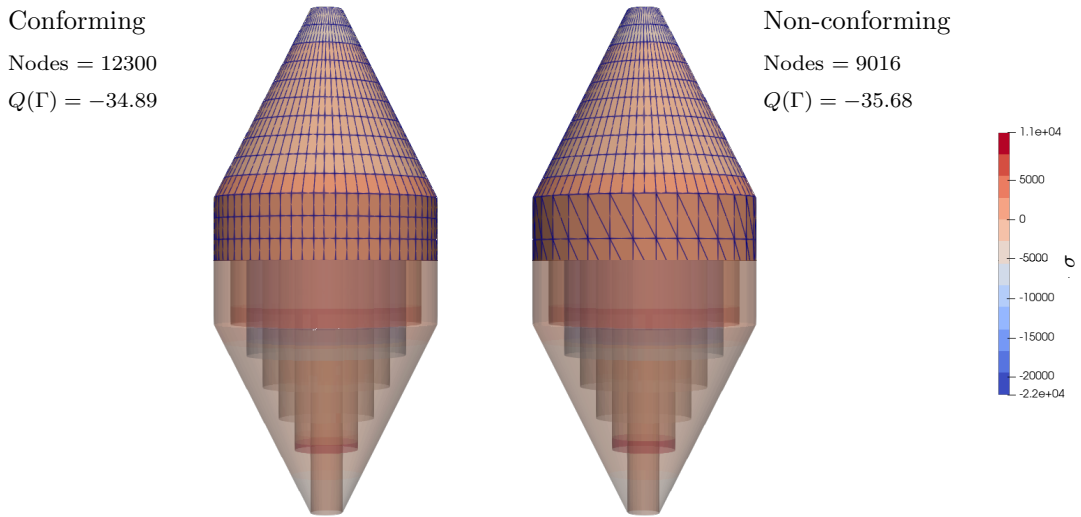}
  \caption{Comparison of the surface charge density obtained by (left) a conforming mesh and (right) a non-conforming one.}
  \label{fig:bushingSol}

\end{figure}
The upper part of \cref{fig:bushingSol} illustrates these two discretizations together with the computed surface charge density $\sigma$. The lower part indicates $\sigma$ along the foils. 
%
The conforming mesh has 12300 nodes and is finer than the non-conforming one with 9016 nodes. There is no visual difference between the results. Looking at the obtained total surface charges $Q(\Gamma)$, the relative difference is $2.2 \cdot 10^{-2}$.

\section{CONCLUSIONS}

The proposed concept minimizes the meshing effort for electrostatic simulations to integrate CAD data more seamlessly into the simulation process. 
We propose a Boundary Element Method (BEM), thereby circumventing the need for volumetric meshing by employing boundary integral equations. Furthermore, the formulation permits discontinuous field quantities, which allows the mesh to be non-conforming. 
As a result, the meshing process can be carried out independently for each trimmed surface of a CAD model, which mitigates meshing problems.
To perform the simulation with non-conforming meshes properly, we enhance the edge data of trimmed CAD models.
This process is done within the CAD system with no need for data translation and with the same accuracy and robustness measures used to set up the model. 
Two numerical experiments demonstrate the quality of the results from the non-conforming BEM simulation. 
%
The current workflow relies on linear elements. In the future, we will investigate the impact of high-order elements on the numerical simulation.

\section{ACKNOWLEDGMENTS}
This work was supported by the Österreichische Forschungsförderungsgesellschaft (FFG), Austria under the grand number: 883886.

\bibliographystyle{plainnat}
\bibliography{eccomas2024}

\begin{thebibliography}{15}
\providecommand{\natexlab}[1]{#1}
\providecommand{\url}[1]{\texttt{#1}}
\expandafter\ifx\csname urlstyle\endcsname\relax
  \providecommand{\doi}[1]{doi: #1}\else
  \providecommand{\doi}{doi: \begingroup \urlstyle{rm}\Url}\fi

\bibitem[Ali et~al.(2016)Ali, Tyacke, Tucker, and Shahpar]{Ali_2016a}
Z.~Ali, J.~Tyacke, P.~G. Tucker, and S.~Shahpar.
\newblock Block topology generation for structured multi-block meshing with
  hierarchical geometry handling.
\newblock \emph{Procedia Engineering}, 163:\penalty0 212--224, 2016.
\newblock \doi{10.1016/j.proeng.2016.11.050}.

\bibitem[Amann et~al.(2014)Amann, Blaszczyk, Of, and Steinbach]{Amann2014a}
Dominic Amann, Andreas Blaszczyk, G{\"u}nther Of, and Olaf Steinbach.
\newblock Simulation of floating potentials in industrial applications by
  boundary element methods.
\newblock \emph{Journal of Mathematics in Industry}, 4\penalty0 (1):\penalty0
  13, 2014.
\newblock ISSN 2190-5983.
\newblock \doi{10.1186/2190-5983-4-13}.

\bibitem[Botsch et~al.(2010)Botsch, Kobbelt, Pauly, Alliez, and
  L{\'e}vy]{botsch2010polygon}
Mario Botsch, Leif Kobbelt, Mark Pauly, Pierre Alliez, and Bruno L{\'e}vy.
\newblock \emph{Polygon mesh processing}.
\newblock AK Peters/CRC Press, 2010.
\newblock \doi{10.1201/b10688}.

\bibitem[Bošnjak et~al.(2023)Bošnjak, Pepe, Schussnig, Schmalstieg, and
  Fries]{Bosnjak_2023a}
D.~Bošnjak, A.~Pepe, R.~Schussnig, D.~Schmalstieg, and T.~P. Fries.
\newblock Higher-order block-structured hex meshing of tubular structures.
\newblock \emph{Engineering with Computers}, 400:\penalty0 1--21, 2023.
\newblock \doi{10.1007/s00366-023-01834-7}.

\bibitem[Bošnjak et~al.(2024)Bošnjak, Pepe, Schussnig, Egger, and
  Fries]{Bosnjak_2024b}
D.~Bošnjak, A.~Pepe, R.~Schussnig, J.~Egger, and T.P. Fries.
\newblock A semi-automatic method for block-structured hexahedral meshing of
  aortic dissections.
\newblock \emph{Int. J. Numer. Method Biomed. Eng.}, 0:\penalty0 accepted,
  2024.
\newblock \doi{10.1002/cnm.3860}.

\bibitem[Erichsen and Sauter(1998)]{erichsen1998}
Stefan Erichsen and Stefan~A Sauter.
\newblock Efficient automatic quadrature in 3-d galerkin bem.
\newblock \emph{Computer methods in applied mechanics and engineering},
  157\penalty0 (3-4):\penalty0 215--224, 1998.

\bibitem[Fogg et~al.(2015)Fogg, Armstrong, and Robinson]{Fogg_2015a}
H.~J. Fogg, C.~G. Armstrong, and T.~T. Robinson.
\newblock Automatic generation of multiblock decompositions of surfaces.
\newblock \emph{International Journal for Numerical Methods in Engineering},
  101:\penalty0 965--991, 2015.
\newblock \doi{10.1002/nme.4825}.

\bibitem[Marussig and Hughes(2018)]{marussig2017a}
Benjamin Marussig and Thomas J.~R. Hughes.
\newblock A review of trimming in isogeometric analysis: Challenges, data
  exchange and simulation aspects.
\newblock \emph{Archives of Computational Methods in Engineering}, 25\penalty0
  (4):\penalty0 1059--1127, 2018.
\newblock ISSN 1886-1784.
\newblock \doi{10.1007/s11831-017-9220-9}.

\bibitem[Massarwi et~al.(2018)Massarwi, van Sosin, and Elber]{Massarwi2018a}
Fady Massarwi, Boris van Sosin, and Gershon Elber.
\newblock Untrimming: Precise conversion of trimmed-surfaces to tensor-product
  surfaces.
\newblock \emph{Computers \& Graphics}, 70:\penalty0 80--91, 2018.
\newblock ISSN 0097-8493.
\newblock \doi{10.1016/j.cag.2017.08.009}.

\bibitem[Owen(1998)]{Owen_1998a}
S.~J. Owen.
\newblock A survey of unstructured mesh generation technology.
\newblock In \emph{International Meshing Roundtable Conference}, 1998.
\newblock \doi{https://api.semanticscholar.org/CorpusID:2675840}.

\bibitem[Sauter and Schwab(2010)]{sauter2010}
Stefan~A Sauter and Christoph Schwab.
\newblock \emph{Boundary element methods}.
\newblock Springer, 2010.

\bibitem[Si(2015)]{Si_2015a}
H.~Si.
\newblock {T}et{G}en, a {D}elaunay-based quality tetrahedral mesh generator.
\newblock \emph{ACM Trans Math Softw}, 41:\penalty0 11, 2015.
\newblock \doi{10.1145/2629697}.

\bibitem[Urick and Marussig(2017)]{urick2017web}
Benjamin Urick and Benjamin Marussig.
\newblock Why cad surface geometry is inexact, 2017.

\bibitem[Urick et~al.(2020)Urick, Crawford, Hughes, Cohen, and
  Riesenfeld]{Urick2020a}
Benjamin Urick, Richard~H. Crawford, Thomas J.~R. Hughes, Elaine Cohen, and
  Richard~F. Riesenfeld.
\newblock Reconstruction of trimmed {NURBS} surfaces for gap-free
  intersections.
\newblock \emph{Journal of Computing and Information Science in Engineering},
  20\penalty0 (5), 07 2020.
\newblock ISSN 1530-9827.
\newblock \doi{10.1115/1.4047427}.
\newblock 051008.

\bibitem[Wassermann et~al.(2019)Wassermann, Kollmannsberger, Yin, Kudela, and
  Rank]{Wassermann2019a}
Benjamin Wassermann, Stefan Kollmannsberger, Shuohui Yin, L\'{a}szl'{o} Kudela,
  and Ernst Rank.
\newblock Integrating {CAD} and numerical analysis: `dirty geometry' handling
  using the finite cell method.
\newblock \emph{Computer Methods in Applied Mechanics and Engineering},
  351:\penalty0 808--835, 2019.
\newblock ISSN 0045-7825.
\newblock \doi{10.1016/j.cma.2019.04.017}.

\end{thebibliography}

\end{document}